%% file: main.tex
\PassOptionsToPackage{table}{xcolor}
\documentclass[11pt,a4paper]{style}
\usepackage{array,booktabs,tabularx,multirow,graphicx,mathtools}
\usepackage{tcolorbox}
\usepackage{fontawesome5}
\tcbuselibrary{breakable}
\usepackage{adjustbox}
\usepackage{placeins}
\usepackage{float}
\input{preamble}

\definecolor{brandorange}{HTML}{ED6000}
\definecolor{absframe}{HTML}{ED6000}
\makeatletter
\DeclareRobustCommand\onedot{\futurelet\@let@token\@onedot}
\def\@onedot{\ifx\@let@token.\else.\null\fi\xspace}

\makeatother

\hypersetup{
  pdftitle={Visual Attribution Limits and Graded Provenance Auditing in Commercial Video-Generation APIs},
  pdfauthor={Haocheng Mei; Yuchuan Tian; Yingte Shu; Chao Xu; Kai Han; Allen Chen; Yunhe Wang}
}

\makeatletter
\renewcommand{\abscontent}{%
  \begin{tcolorbox}[breakable,colframe=absframe,colback=absframe!5,boxrule=1pt,arc=8pt,
    left=10pt,right=10pt,top=10pt,bottom=10pt]
  \centerline{\fontsize{15pt}{14pt}\selectfont\textbf{Abstract}}\vspace{2ex}
  {\absfont\theabstract}
  \par\smallskip
  {\centering\hypersetup{urlcolor=blue}%
    \faGithub\ \textbf{GitHub:} \url{https://github.com/TokenRhythm/videoapi}\par}
  \end{tcolorbox}%
}
\renewcommand{\maketitle}{%
  \begingroup\setlength{\parindent}{0pt}
  \vspace*{3pt}
  {\centering\titlefont\@title\par}
  \vskip6pt
  {\centering\normalfont\fontsize{12}{15}\selectfont TokenRhythm Technologies\par}
  \vskip9pt
  \endgroup
  \thispagestyle{firststyle}\abscontent
}
\makeatother

\fancypagestyle{firststyle}{%
  \fancyhf{}
  \fancyhead[L]{\smash{\raisebox{0.22em}{\includegraphics[height=1.25em]{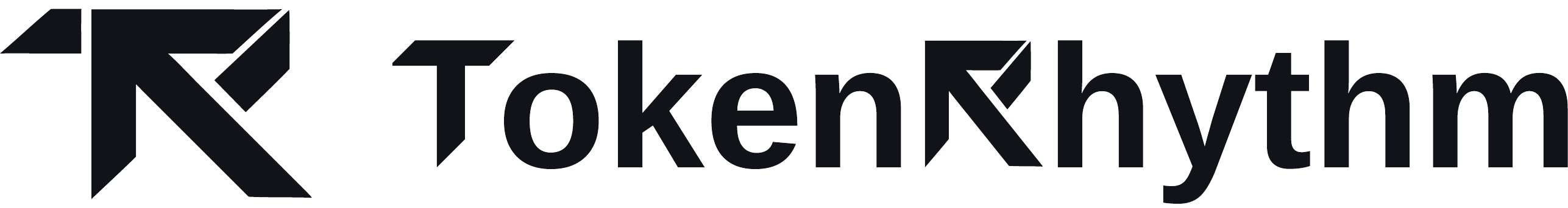}}}}
  \fancyhead[R]{}

}

\title{Visual Attribution Limits and Graded Provenance Auditing in~Commercial Video-Generation APIs}
\author{}
\date{}

\input{sections/00_abstract}

\begin{document}
\maketitle
\input{sections/01_intro}
\input{sections/02_related}
\input{sections/03_problem}
\input{sections/04_method}
\input{sections/05_experiments}
\input{sections/06_limitations}
\input{sections/07_conclusion}
\section*{Contributors}
\noindent\begin{minipage}{\linewidth}
  \raggedright
  \mbox{Haocheng Mei\textsuperscript{1}}, \mbox{Yuchuan Tian\textsuperscript{1}}, \mbox{Yingte Shu\textsuperscript{1}}, \mbox{Chao Xu\textsuperscript{1}},\\
  \mbox{Kai Han\textsuperscript{2}}, \mbox{Allen Chen\textsuperscript{3}}, \mbox{Yunhe Wang\textsuperscript{2}}\par\smallskip
  {\small \mbox{\textsuperscript{1}Peking University}\quad \mbox{\textsuperscript{2}TokenRhythm Technologies}\quad \mbox{\textsuperscript{3}Tsinghua University}\par}
\end{minipage}
\bibliographystyle{ieeenat_fullname}
\bibliography{refs}
\clearpage
\appendix
\input{sections/09_appendix}
\end{document}

%% file: preamble.tex
\usepackage{graphicx}
\usepackage{amsmath,amssymb}
\usepackage{booktabs}
\usepackage{array} %
\usepackage{multirow}
\usepackage{xspace}
\usepackage{pifont}
\usepackage[dvipsnames]{xcolor}

\newcommand{\prelim}[1]{{\color{BurntOrange}#1}}

\renewcommand{\prelim}[1]{#1} %

\DeclareMathOperator{\Bin}{Bin}

%% file: sections/00_abstract.tex
\begin{abstract}
Commercial video-generation APIs bill by model and tier, while intermediaries could substitute cheaper models, lower tiers, or cached clips. %
What can a delivered file establish about its origin?
We combine measurements of passive attribution and cryptographic provenance into a graded audit.
We test attribution probes and zero-shot detectors under re-encoding, downscaling, and frame-rate changes, and survey C2PA and GB~45438 metadata across 19 commercial endpoints. %
Only authenticated provenance supports accusations: tier contradictions or duplicate signed instances under fresh-generation contracts, excluding retries and idempotent returns. %
Unverifiable signed-endpoint deliveries trigger original-file demands under pass-through clauses; archive-reuse and delivery-parameter screens support review only. %
Six-model container and bitstream accuracy falls from 83.9\% on raw files to 45.3\% after uniform re-encoding, using different feature sets. %
Subsampling real references from 24 to 3~fps raises pooled D3 AUROC from 0.845 to 0.988. %
Only BytePlus authenticates tier claims under our pinned policy, including 187/187 audited deliveries across three Seedance tiers; tested remuxing and re-encoding leave processed Seedance files unverifiable. %
The audit identifies authenticated contradictions in 100/100 raw 720p signed tier-swap cases, with zero observed false flags on the tested honest pools. %
\end{abstract}

%% file: sections/01_intro.tex
\section{Introduction}\label{sec:intro}

Commercial video-generation APIs price each generated clip by the requested model and tier, ranging from \$0.13 to \$1.17 per delivery across the 19 endpoints in our pilot ledger. %
Within individual model developers, lower-priced tiers cost 25--79\% less than premium tiers across four vendors we compared. %
When requests pass through a third-party gateway, platform, or reseller, that price spread creates a potential incentive to serve a cheaper model or tier, reduce rendering compute, substitute a fraction of jobs, or return a cached clip (threats S1--S5 in Sec.~\ref{sec:problem}). %
Yet the buyer receives only the returned video file, timing and billing records, and final gateway job objects that, in our inspections, expose no upstream model or tier identity. %
Seeded replay offers no independent baseline either: byte-identical seeded requests sent roughly 3~h apart failed a pre-registered 0.90 SSIM gate on all six tested endpoints, peaking at 0.876 across three pairs per endpoint. %
Together, these constraints raise a basic forensic problem: determining what a delivered video alone can establish about the model and tier that produced it.

\begin{figure*}[t]
    \centering
    \includegraphics[width=\textwidth]{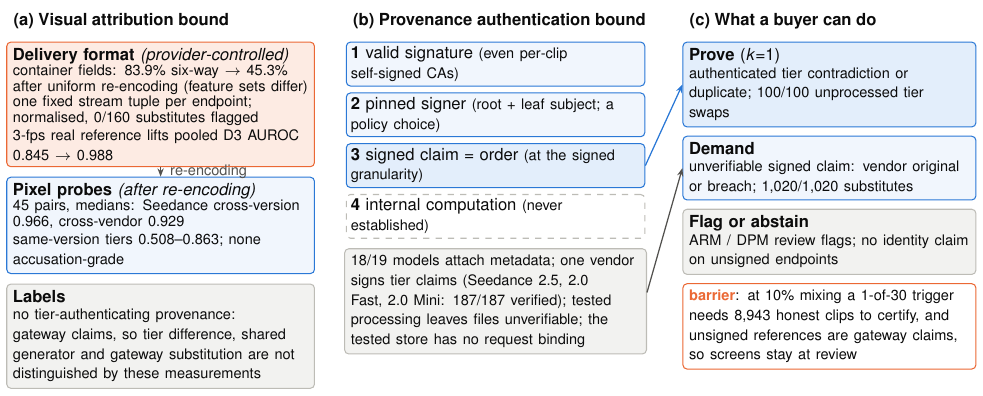}
    \caption{\textbf{The attribution--authentication gap in commercial video-generation APIs.} (a)~\emph{Visual attribution bound:} provider-controlled delivery formatting and reference frame-rate choices shift raw attribution and zero-shot detection (six-way container and bitstream accuracy falls from 83.9\% raw to 45.3\% after uniform re-encoding as available features narrow), while pixel probes after uniform re-encoding separate most versions and vendors yet remain below accusation grade, and unauthenticated gateway labels cannot separate tier differences from a shared generator or substitution. (b)~\emph{Provenance authentication bound:} signature validity, pinned signer trust, signed claim granularity, and unverified internal computation form four distinct levels; 18 of 19 surveyed models attach metadata, but only one vendor authenticates tiers under our pinned policy, tested processing leaves C2PA files unverifiable, and the inspected store lacks request binding. (c)~\emph{Buyer recourse:} because statistical triggers face steep honest-clip certification costs and unauthenticated gateway references, recourse divides into \emph{Prove}, \emph{Demand}, and \emph{Flag or abstain} via review-grade ARM and DPM screens.} %
    \label{fig:overview}
\end{figure*}

Our measurements show that file-level evidence is bounded from two directions, leaving a wide gap between passive visual forensics and cryptographic provenance (Fig.~\ref{fig:overview}).
Six-way endpoint classification reached 83.9\% using container and bitstream features on raw files and 45.3\% using bitstream features after uniform H.264 re-encoding. %
Even inside nominal 24-fps H.264 encodes, subsampling a 24-fps real reference cohort to 3~fps raises pooled AUROC of the zero-shot detector D3~\citep{d3video2025} from 0.845 to 0.988 while pulling STALL~\citep{stall2025} downward, flipping their post-hoc ranking on the same real clips. %
Once deliveries are uniformly re-encoded, pixel probes retrained within condition still separate most vendors and model versions across 45 pairs, but performance weakens between tiers of the same version, spanning near chance on Wan~3.0~Prime vs.\ 3.0 (0.508--0.551 AUROC) to 0.863 on C2PA-authenticated Seedance~2.0~Mini vs.\ 2.0~Fast. %
None of these passive probes is reliable enough to support an accusation, and on endpoints without tier-authenticating provenance under our pinned policy, gateway labels cannot independently establish the true generating tier. %

Cryptographic provenance carries a complementary set of limits.
A valid C2PA signature~\citep{c2paspec} authenticates only a trusted signer's claim about the original file at the signed granularity, never internal weights or compute. %
Across 19 surveyed models spanning 605 deliveries (\prelim{nine sampled at $n = 1$}), 18 attach metadata, yet C2PA under our pinned signer policy authenticates tier claims for only BytePlus across all 187 audited deliveries from three multi-clip Seedance~2.x tiers (\prelim{plus Seedance~2.0 at $n = 1$}). %
Among the remaining endpoints, Grok signs a shared product name across tiers using per-clip self-signed certificate authorities (CAs), \prelim{single-clip checks verify FLUX.3 only at model level and fail credential validation on Runway Gen-4.5 despite a C2PA-listed root ($n = 1$ each)}, and ten models attach vendor-level GB~45438-2025~\citep{gb45438} metadata whose integrity fields went untested without vendor-supplied schemes and keys. %
Routine remuxing and re-encoding also leave previously verifiable C2PA files unverifiable by stripping their manifests in \prelim{preliminary cross-vendor checks ($n \le 3$ clips per vendor and transform)}, and the inspected Seedance manifest store omits both request binding and trusted signing timestamps. %

To bridge this gap, we study model-substitution forensics by jointly evaluating passive attribution, deployed provenance, and the practical recourse open to a buyer.
In a 30-clip audit under 10\% partial substitution, certifying a single-clip statistical trigger requires 8,943 zero-failure honest calibration clips per endpoint, while multi-clip tests against wholesale substitution still face unauthenticated gateway references, unseen substitute models, and unestablished stationarity across sessions. %
A buyer can prove a tier contradiction by comparing one authenticated delivery with its order. To establish duplicate billing, its signed instance ID must match a previous authenticated delivery under a separately billed fresh-generation order, after retries and idempotent returns are excluded. An unverifiable delivery from a signed endpoint triggers a demand for the vendor original under a pass-through clause. Elsewhere, archive reuse matching (ARM)~\citep{phash2010,ssim2004} and delivery-parameter matching (DPM) can trigger review. %
Our main contributions are threefold:

\begin{itemize}
    \item \textbf{Passive visual forensics bounds:} Across two 8-fps zero-shot detectors (D3 and STALL), a 24-fps real reference cohort, and 10 commercial models (45 pairs), we extend the frame-duplication analysis of \citet{stall2025} with matched-content interventions showing that the relative ranking of D3 and STALL can reverse inside nominal 24-fps streams, while re-encoded visual attribution depends on processing and degrades from strong cross-vendor separation to near-chance same-version tiers. %
    \item \textbf{Deployed provenance census:} Across a 19-model, 605-delivery corpus spanning C2PA and GB~45438-2025, we build a pinned-policy verifier validated on 18 adversarial controls to map metadata coverage, signer trust, claim granularity, transcoding fragility, and missing request binding. %
    \item \textbf{Provenance-first graded audit:} Grounded in calibration and reference-authenticity bounds that keep passive classifiers out of accusation triggers, we formulate and evaluate \emph{Prove}, \emph{Demand}, and \emph{Flag or abstain} rules across constructed substitutions, parameter-normalised substitutes, and archive-reuse deliveries. %
\end{itemize}

%% file: sections/02_related.tex
\section{Related Work}\label{sec:related}

\paragraph{Generated-video detection.}
GenVideo and GenVidBench provide benchmarks for real-versus-synthetic video detection~\citep{demamba2024,genvidbench2025}. Binary video detectors include learned classifiers~\citep{vahdati2024} and training-free methods based on second-order temporal differences (D3)~\citep{d3video2025} and spatial-temporal likelihoods (STALL)~\citep{stall2025}.
\citet{stall2025} traced inflated D3 scores on GenVideo's MSR-VTT split to duplication of frames from 3-fps real videos. They also reported stable STALL performance when both classes were uniformly subsampled. We test both detectors against the same commercial deliveries, varying the effective rate of the real clips inside nominal 24-fps encodes and adding a matched 3-fps control for both classes. %

\paragraph{Visual attribution and classical forensics.}
Earlier studies attributed images and videos to generators using source fingerprints~\citep{marra2019gans,yu2019attributing,saga2026}. General-purpose encoders such as SigLIP, DINOv3, and X-CLIP provide visual representations~\citep{siglip2023,dinov3_2025,xclip2022}. Our setting includes commercial endpoints from model families described in the Wan and Seedance reports~\citep{wan2025,seedance2025}. Classical forensics studies spatial resampling~\citep{popescu2005resampling,kirchner2008resampling}, frame-rate conversion~\citep{bian2014fruc,yoon2021fruc}, and double compression~\citep{wang2006double}. Video similarity learning~\citep{visil2019}, perceptual image hashing~\citep{phash2010}, and structural similarity~\citep{ssim2004} offer complementary ways to compare content.
Closed attribution benchmarks typically assume direct access to generator outputs and ground-truth labels.
In commercial APIs, by contrast, intermediaries control container muxing and re-encoding, may apply super-resolution~\citep{realesrgan2021} or frame interpolation~\citep{rife2022,film2022}, and expose unverified gateway labels whenever tier-level provenance is absent, leaving re-encoded tier separability pair-specific and below accusation grade. %

\paragraph{Provenance, watermarking, and regulation.}
C2PA~\citep{c2paspec} cryptographically binds signed assertions to asset hashes, though recent work highlights trust-list and specification gaps~\citep{c2pafall2026} as well as manifest--watermark desynchronisation~\citep{authcontra2026}.
Invisible watermarks have been developed for generated images~\citep{stablesig2023,treering2023} and videos~\citep{videoseal2024,videoshield2025}. Work on image watermarks has tested their robustness and shown that regeneration can remove them~\citep{zhao2024removable,waves2024}. China's GB~45438-2025 standard specifies metadata for production and distribution~\citep{gb45438}. The accompanying measures assign labeling duties to generation and dissemination services in the circumstances they cover~\citep{cacmeasures2025}.
Rather than proposing a new watermark, we examine what deployed C2PA and GB~45438 credentials actually establish for an API buyer across signer trust, claim granularity, transcoding survival, and request binding. %

\paragraph{Auditing model substitution in APIs.}
FARE learns an acceptance region from images sampled from an enrolled generator and tests whether subsequent images are consistent with it~\citep{fare2026}. In our video setting, an unsigned gateway may also supply the reference labels, leaving the identity of the enrolled tier unverified. Audits of text-generation APIs test output distribution equality~\citep{mettest2025}, cheaper-model substitution~\citep{cai2025paying}, misrepresented models in shadow gateways~\citep{realmoney2026}, diluted routing~\citep{iris2026}, anonymous model identity~\citep{fourstage2026}, and multi-hop attestation~\citep{aex2026}.
In our video audit, each additional reference sample requires a paid clip generation. The gateway provides no signed request--response attestation, and on unsigned endpoints it supplies the only labels for the reference clips. At the target false-accusation rate in Sec.~\ref{sec:audit_rules}, calibrating a single-clip trigger would require thousands of honest clips per endpoint. %
These constraints motivate a graded audit anchored on deployed vendor signatures rather than standalone statistical classifiers. %

%% file: sections/03_problem.tex
\section{API Delivery Setting and Commercial Corpus}\label{sec:problem}

\paragraph{Parties and observations.}
In our setting, a \emph{buyer} (auditor) requests clips from a named model and tier (\emph{endpoint}) through an intermediary \emph{provider} (a platform, gateway, or reseller), while the upstream \emph{vendor} trains the generator and, where supported, signs provenance.
Each returned file constitutes a \emph{delivery}, and an \emph{audit} evaluates $n = 30$ deliveries from a single endpoint. %
Per request, the buyer specifies the prompt, seed, target resolution, aspect ratio, duration, image-to-video (I2V) first frame, audio flag, and submission time, and observes only the raw returned bytes, gateway response fields, timestamps, and billed cost. %
The final gateway job records we inspected carry only \texttt{id}, \texttt{generation\_id}, \texttt{status}, \texttt{usage.cost}, \texttt{usage.is\_byok}, and \texttt{expires\_at}, revealing no upstream vendor or tier identity; explicit pixel dimensions, last-frame conditioning, and multi-account probing were not tested. %

\paragraph{Replay limits.}
Seeded requests also fail to reproduce through the gateway, preventing a buyer from building deterministic reference clips by replay. %
Across six tested endpoints (three pairs of byte-identical seeded requests per endpoint submitted 176.8--189.9~min apart), 0/6 endpoints and 0/18 pairs cleared a pre-registered 0.90 SSIM gate: similarity peaked at 0.876 on Wan~3.0, while per-endpoint minima fell between 0.208 and 0.474 (direct vendor-API determinism was not tested). %

\paragraph{Substitution threats and scope boundaries.}
A provider that controls the delivery pipeline can substitute outputs in five ways: routing to a cross-vendor model (S1), routing to a cheaper within-vendor tier (S2), rendering the claimed model at reduced compute (S3) via spatial upscaling (S3-sp), temporal interpolation (S3-t), or fewer denoising steps (S3-st), mixing substitutes into a fraction $p$ of jobs (S4), or serving cached clips from earlier jobs (S5).
Along the way, the provider may re-encode video, strip or transplant metadata, apply super-resolution~\citep{realesrgan2021} or frame interpolation~\citep{rife2022,film2022}, or route selectively.
On endpoints with tier-authenticating provenance, post-generation S3 re-rendering breaks C2PA asset binding and triggers an original-file demand, whereas passive visual detection of unsigned S3 compute downgrades was not evaluated. %
Compute degradation performed by the vendor prior to signing, signing-key compromise, and legal enforcement lie outside our scope, since cryptographic signatures attest to vendor claims rather than internal execution. %

\begin{table*}[t]
    \centering
    \caption{\textbf{Commercial video corpus and offline evaluation sets.} \emph{Main}, \emph{Census}, and \emph{Low-res} form our 605-delivery purchased corpus across 19 models (\prelim{9 single-clip \emph{Census}-only models at $n = 1$}); \prelim{the second-session re-check is a separate 6-clip purchase ($n = 2$ sessions)}, and \emph{Substitution} constructs offline swap cases. New clip counts include retained purchases only.} %
    \label{tab:datasets}
    \footnotesize
    \setlength{\tabcolsep}{2.5pt}
    \begin{tabularx}{\linewidth}{@{}l>{\raggedright\arraybackslash}Xr>{\raggedright\arraybackslash}p{0.075\linewidth}>{\raggedright\arraybackslash}p{0.112\linewidth}r>{\raggedright\arraybackslash}p{0.175\linewidth}@{}}
        \toprule
        Set & Contents and collection protocol & \shortstack{New\\clips} & Models & Target spec & Cost & Primary role \\
        \midrule
        \emph{Main} & 536 unseeded requests, 528 completions, 520 kept (8 from 1 refused prompt set aside); 6 models $\times$ 60 prompts + 4 $\times$ 40 shared prompts & 520 & 10 & 720p, 16:9, 5\,s & \$270.86 & Detection, 45-pair separability, provenance \\ %
        \emph{Census} & 1 clip per model on 1 neutral prompt (3 HTTP~403 Veo~3.1 ToS refusals excluded); Hailuo~2.3 at 1080p 6\,s, 3 at 2K, 3-max at 768p & 19 & 19 & 720p, 16:9, 5\,s & \$10.24 & 19-model provenance \& pricing survey \\ %
        \emph{Low-res} & Pre-registered validation collection (thresholds fixed before purchase); includes seeded replay and regeneration pairs & 66 & 7 & 480p & \$13.33 & Replay, C2PA, ARM \& 480p validation \\ %
        \midrule
        \textbf{Corpus total} & Combined first-collection originals (\prelim{9 \emph{Census}-only models have $n = 1$ clip}) & \textbf{605} & \textbf{19} & --- & \textbf{\$294.43} & Full re-extraction; ARM negatives use the 586-clip \emph{Main} + \emph{Low-res} subset \\ %
        \midrule
        \prelim{\emph{Re-check} (prelim.)} & \prelim{Separate second session 8 days later across 6 single-clip \emph{Census} models (1 clip each, $n = 2$ sessions)} & \prelim{6} & \prelim{6} & \prelim{Varies} & \prelim{\$2.84} & \prelim{Cross-session regime \& DPM check} \\ %
        \emph{Substitution} & Offline prompt-aligned pairing of \emph{Main} and \emph{Low-res}: 2,061 fresh-sub.\ cases + 309 honest-pool entries (209 distinct clips, \prelim{incl.\ 9 at 480p}) & 0 & 6 pairs & 720p; \prelim{480p} & \$0.00 & Graded audit \& 30-clip simulations \\ %
        \bottomrule
    \end{tabularx}
\end{table*}

\paragraph{Commercial video corpus and evaluation sets.}
All commercial deliveries were purchased through OpenRouter under the collection protocols and budgets summarised in Tab.~\ref{tab:datasets}. %
Thresholds for the \emph{Low-res} validation experiments were pre-registered prior to purchase, while subsequent provenance and visual analyses ran frozen evaluation procedures on the collected corpus; we identify post-hoc analyses and proposed rules in the text. The planned public release includes research scripts, prompts, verifier policies, and redacted aggregate results. It excludes purchased media, billing records, and some control-generation scripts. %

Within the 605-clip primary corpus, \emph{Main} supplies 60 clips across 60 prompts each for Seedance~2.5, 2.0~Mini, 1.5~Pro, Grok~Imagine~1.5, Grok~1.5~Lite, and HappyHorse~1.1, and 40 clips on a shared 40-prompt subset each for Seedance~2.0~Fast, Kling~3.0~Std, Wan~3.0, and Wan~3.0~Prime (after replacing one provider-refused prompt under our frozen reserve rule and setting aside its eight completed clips). %
\emph{Census} samples all 19 accessible endpoints on a single neutral prompt, excluding three Google Veo~3.1 endpoints that returned HTTP~403 Terms-of-Service refusals without workaround, while \emph{Low-res} (distinct from the D3 detector below) collects 480p validation clips across seven endpoints, leaving \prelim{nine \emph{Census}-only models at $n = 1$}. %
\prelim{Eight days after the main collection, we re-purchased one clip each from six single-clip \emph{Census} models (FLUX.3, Seedance~2.0, Kling~3.0~Pro, Hailuo~3-max, Wan~2.6, and Wan~2.7) as a preliminary cross-session check (\emph{Re-check}, $n = 2$ sessions).} %
Finally, \emph{Substitution} pairs \emph{Main} and \emph{Low-res} clips on shared prompts for six ordered model pairs across seven pair--resolution combinations and their processing variants to build 2,061 offline ``fresh-substitution'' cases (denoting prompt-paired substitutes as opposed to archive reuse, rather than new API spend) alongside 309 honest-pool entries drawn from 209 distinct clips (\prelim{including $n = 9$ preliminary clips at 480p}). %

\paragraph{Processing conditions.}
Our visual experiments compare raw deliveries against three standardised processing stages. %
\textbf{N1} remuxes the original video stream into a clean container with metadata and audio stripped. %
\textbf{N2} uniformly re-encodes video as H.264 at 720p, nominal 24~fps, and constant rate factor (CRF) 23, again stripping metadata and audio. %
The \textbf{480p CRF-32 round trip} downscales clips to 480p and re-encodes at CRF~32 to probe heavier lossy compression. %

%% file: sections/04_method.tex
\section{What Passive Visual Forensics Establishes}\label{sec:visual}

\begin{figure*}[t]
    \centering
    \includegraphics[width=\textwidth]{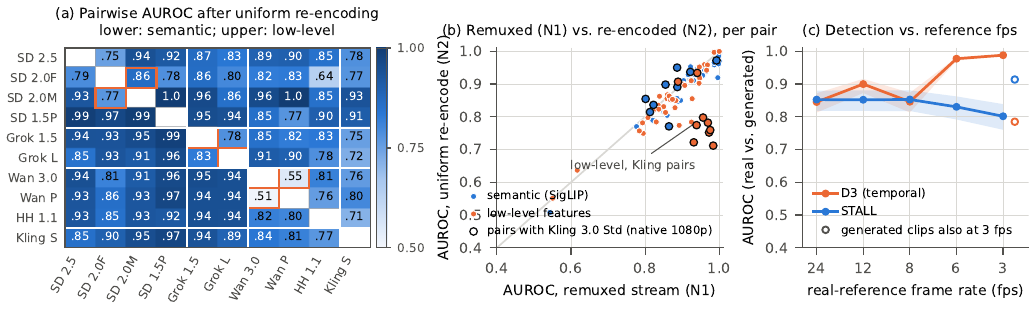}
    \caption{\textbf{Pairwise separability, processing shifts, and the reference frame-rate confound.} (a)~N2 AUROC (720p, 24~fps, CRF~23) across all 45 \emph{Main} pairs for semantic SigLIP (lower) and low-level (upper) probes; white lines divide vendors and orange borders mark same-version tiers. (b)~Per-pair AUROC at N1 (remuxed video, metadata and audio stripped) vs.\ N2 for semantic (blue) and low-level (orange) features; black rings mark the nine native-1080p Kling~3.0~Std pairs (8 of 9 low-level points lie below the diagonal, whereas semantic points stay near it). (c)~Zero-shot D3 detector and STALL AUROC across 520 \emph{Main} clips (bands: 10-model min--max) vs.\ the same 1,000 GenVideo real clips subsampled to $24, 12, 8, 6, 3$~fps inside nominal 24-fps N2 encoding (equally spaced axis): 8~fps matches 24~fps ($+0.001$ D3, $+0.000$ STALL), while post-hoc ranking in this shared real cohort flips from D3$>$STALL on 3/10 models (24, 8~fps) to 10/10 (12, 6, 3~fps) and 0/10 when generated clips are also 3~fps (hollow). Abbreviations: SD = Seedance (2.0F = 2.0~Fast, 2.0M = 2.0~Mini, 1.5P = 1.5~Pro), Grok~L = Grok~1.5~Lite, Wan~P = Wan~3.0~Prime, HH = HappyHorse, and Kling~S = Kling~3.0~Std.} %
    \label{fig:visual_findings}
\end{figure*}

We begin by evaluating two training-free generated-video detectors that sample decoded frames at 8~fps: the D3 detector~\citep{d3video2025}, based on second-order temporal feature differences, and STALL~\citep{stall2025}, which combines spatial and temporal likelihoods. %
We evaluate both detectors on 1,000 24-fps real clips and all 520 \emph{Main} deliveries. For supervised attribution, we train separate classifiers for the 45 pairs of \emph{Main} models, using only the generated clips from prompts shared by each pair (Fig.~\ref{fig:visual_findings}). %
Two overarching patterns emerge: effective temporal sampling rather than H.264 compression drives large shifts in zero-shot detection even when every file shares nominal 24-fps encoding, and pairwise visual attribution varies sharply with both processing stage and model pair. %

\subsection{Temporal Sampling vs.\ Compression in Zero-Shot Detection}\label{sec:visual_detection}

\begin{table*}[t]
    \centering
    \caption{\textbf{Controlled frame-rate and compression interventions on zero-shot detection.} Pooled AUROC ($P(\text{real} > \text{generated})$) and pooled shift $\Delta$ [95\% CI] vs.\ the 24-fps CRF-23 baseline across the same 1,000 GenVideo real clips and 520 \emph{Main} deliveries encoded at nominal 24~fps in N2. The rightmost column (descriptive, post-hoc) counts the 10 \emph{Main} models where D3 exceeds STALL in this shared real cohort.} %
    \label{tab:temporal_sampling}
    \footnotesize
    \setlength{\tabcolsep}{2.2pt}
    \adjustbox{max width=\linewidth}{%
\begin{tabular}{@{}lccccc@{}}
        \toprule
        Intervention (in 24-fps N2) & D3 & D3 $\Delta$ [95\% CI] & STALL & STALL $\Delta$ [95\% CI] & D3$>$ST. \\ %
        \midrule
        24\,fps real, CRF~23 (base) & 0.845 & --- & 0.852 & --- & 3/10 \\ %
        24\,fps real, CRF~18 & 0.845 & $-0.000$ $[-0.003, 0.003]$ & 0.853 & $+0.001$ $[-0.002, 0.003]$ & --- \\ %
        24\,fps real, CRF~28 & 0.846 & $+0.001$ $[-0.002, 0.004]$ & 0.854 & $+0.001$ $[-0.001, 0.004]$ & --- \\ %
        \midrule
        12\,fps real vs.\ 24\,fps gen. & 0.900 & $+0.055$ $[0.048, 0.062]$ & 0.852 & $-0.000$ $[-0.003, 0.003]$ & 10/10 \\ %
        8\,fps real vs.\ 24\,fps gen. & 0.846 & $+0.001$ $[-0.002, 0.004]$ & 0.853 & $+0.000$ $[-0.003, 0.004]$ & 3/10 \\ %
        6\,fps real vs.\ 24\,fps gen. & 0.977 & $+0.132$ $[0.116, 0.149]$ & 0.830 & $-0.022$ $[-0.027, -0.017]$ & 10/10 \\ %
        3\,fps real vs.\ 24\,fps gen. & 0.988 & $+0.143$ $[0.126, 0.161]$ & 0.801 & $-0.051$ $[-0.058, -0.044]$ & 10/10 \\ %
        \midrule
        3\,fps real vs.\ 3\,fps gen. & 0.785 & $-0.060$ $[-0.076, -0.044]$ & 0.914 & $+0.061$ $[0.050, 0.074]$ & 0/10 \\ %
        \bottomrule
    \end{tabular}%
}
\end{table*}

Even when real and synthetic videos are packaged identically as nominal 24-fps H.264 streams, the effective frame rate of the real reference clips pushes D3 and STALL in opposite directions (Fig.~\ref{fig:visual_findings}c, Tab.~\ref{tab:temporal_sampling}). %
\citet[Appendix~A.4.2]{stall2025} documented this sampling artifact on MSR-VTT. With 1,000 3-fps MSR-VTT clips from GenVideo-Val as our real references, D3 reached 0.989--0.992 AUROC across the 10 \emph{Main} models while STALL stayed at 0.780--0.853, confounding frame rate with source content. %
Replacing that reference set with 1,000 24-fps GenVideo real clips~\citep{demamba2024} brought the two detectors into close agreement (0.813--0.872 per model for D3 and 0.818--0.880 for STALL). %
To isolate temporal sampling from scene content, we then subsampled those same 1,000 GenVideo clips to $f \in \{24, 12, 8, 6, 3\}$~fps before encoding them at nominal 24~fps in N2; our pipeline sanity check reproduced all 1,000 frozen 24-fps encodes byte-for-byte and matched baseline scores within $10^{-6}$. %

As Tab.~\ref{tab:temporal_sampling} shows, varying H.264 compression quality alone (CRF~18 or 28 at 24~fps) changes pooled AUROC by less than 0.002 for either detector. %
Subsampling the real references to 3~fps, by contrast, lifts pooled D3 AUROC from 0.845 to 0.988 ($+0.143$) while lowering STALL from 0.852 to 0.801 ($-0.051$), violating STALL's pre-registered $<0.05$ invariance threshold. %
Across intermediate rates, D3 climbs at 12~fps, drops back to baseline at 8~fps, and nears ceiling separation at 6 and 3~fps. %
Applying the symmetric 3-fps subsampling control to the generated clips as well pulls D3 below baseline to 0.785 ($-0.060$) and lifts STALL to 0.914, satisfying both pre-registered causal criteria for D3 (a real-only 3-fps gain of at least 0.05 with confidence interval above zero, and a symmetric-control shift below half that excess). %
In a post-hoc comparison across individual models on this real cohort, the relative detector ranking flips accordingly: D3 outperforms STALL on only 3/10 models at 24 and 8~fps, on all 10/10 models at 12, 6, and 3~fps, and on 0/10 models under the symmetric 3-fps control. %

This non-monotonic curve is consistent with how an 8-fps detector stride lands on duplicated frames inside a nominal 24-fps stream: %
\begin{itemize}
    \item at 12~fps duplicated to 24~fps, an 8-fps stride alternates between one- and two-frame source steps; %
    \item at 8~fps encoded at 24~fps, an 8-fps stride lands cleanly on every unique frame, matching native 24-fps sampling; %
    \item at 6 and 3~fps encoded at 24~fps, consecutive 8-fps samples repeatedly land on identical duplicated frames. %
\end{itemize}
Inspecting cached frozen features post hoc confirms that D3's AUROC closely tracks the standard deviation of its second-order feature differences (within 0.004 AUROC across all six conditions). %
On the real cohort, that second-order spread rises from a median of 3.440 at 24~fps to 4.068 at 12~fps even as mean first-order step size barely changes (8.781 vs.\ 8.582); it returns to 3.447 at 8~fps, then jumps to 7.066 and 9.124 at 6 and 3~fps as repeated frames depress first-order differences to 7.403 and 4.751. %
STALL's sensitivity instead comes almost entirely from its temporal likelihood term, whose standalone AUROC falls from 0.791 at 24~fps to 0.715 at 3~fps and rebounds to 0.923 under the symmetric control while its spatial term stays narrow (0.846--0.858); causal mediation via frame deduplication prior to scoring remains untested. %

\subsection{Packaging Ladder and Stream-Tuple Shortcuts}\label{sec:visual_format}

\begin{table}[t]
    \centering
    \caption{\textbf{Container and bitstream classification ladder across the six 60-clip \emph{Main} models} (Seedance~2.5, 2.0~Mini, 1.5~Pro, Grok~1.5, Grok~1.5~Lite, and HappyHorse~1.1; 16.7\% chance). Gradient-boosted classifiers are refit within each step, and feature sets narrow across steps.} %
    \label{tab:packaging_ladder}
    \footnotesize
    \setlength{\tabcolsep}{3pt}
    \begin{tabular}{@{}l>{\raggedright\arraybackslash}p{0.47\columnwidth}rc@{}}
        \toprule
        Files & Feature set & Clips & Six-way acc. \\
        \midrule
        Raw & All container and bitstream fields & 360 & 83.9\% \\ %
        Raw & Container and bitstream, provenance removed & 360 & 83.3\% \\ %
        N1 & Remuxed container and bitstream, no provenance & 360 & 77.5\% \\ %
        N2 & Bitstream statistics only (CRF-23 re-encode) & 360 & 45.3\% \\ %
        \bottomrule
    \end{tabular}
\end{table}

Raw API files also carry endpoint-specific packaging information.
Across the six 60-clip \emph{Main} models (360 clips), a gradient-boosted classifier reaches 83.9\% six-way accuracy on raw files, remains high after stripping provenance tags and remuxing into N1, and drops to 45.3\% (0.452 macro-F1 vs.\ 16.7\% chance) after uniform N2 re-encoding (Tab.~\ref{tab:packaging_ladder}). %
At N2 only bitstream statistics survive (bit rates, file and packet sizes, and I/P/B frame counts), meaning both media processing and available features change along the ladder; whether the residual 45.3\% reflects content-driven rate adaptation by the H.264 encoder was not tested separately. %

\begin{table}[t]
    \centering
    \caption{\textbf{Default nine-field stream-parameter tuples on the four claimed 720p DPM endpoints in \emph{Main}.} In each training fold, every reference delivery shares one identical nine-field tuple (48/48 per fold on Seedance~2.5 and Grok~1.5; 32/32 on Wan~3.0~Prime and Kling~3.0~Std); $d$ denotes audio-minus-video stream duration.} %
    \label{tab:tuples}
    \footnotesize
    \setlength{\tabcolsep}{2.2pt}
    \begin{tabular}{@{}>{\raggedright\arraybackslash}p{0.25\columnwidth}>{\raggedright\arraybackslash}p{0.28\columnwidth}>{\raggedright\arraybackslash}p{0.27\columnwidth}>{\raggedright\arraybackslash}p{0.14\columnwidth}@{}}
        \toprule
        Pair (claimed $\leftarrow$ sub.) & Claimed tuple & Substitute tuple & Diff.\ fields \\
        \midrule
        Kling~3.0~Std $\leftarrow$ Wan~3.0 & $1920{\times}1080$, 24\,fps, 121f, no audio & $1280{\times}720$, 30\,fps, 150f, AAC 44.1\,kHz, 2\,ch, $d{=}+38{,}005\,\mu$s & Every field \\ %
        Seedance~2.5 $\leftarrow$ HH~1.1 & $1280{\times}720$, 24\,fps, 121f, no audio & Same video; AAC 24\,kHz, 2\,ch, $d{=}+121{,}000\,\mu$s & Audio track \\ %
        Grok~1.5 $\leftarrow$ 1.5~Lite & $1280{\times}720$, 24\,fps, 121f, AAC 48\,kHz, 2\,ch, $d{=}-41{,}667\,\mu$s & Identical except $d{=}-667\,\mu$s & Offset $d$ only \\ %
        Wan~Prime $\leftarrow$ Wan~3.0 & $1280{\times}720$, 30\,fps, 150f, AAC 44.1\,kHz, 2\,ch, $d{=}+38{,}005\,\mu$s & Identical & None \\ %
        Seedance~2.5 $\leftarrow$ Mini/Fast & $1280{\times}720$, 24\,fps, 121f, no audio & Identical (720p) & None \\ %
        \bottomrule
    \end{tabular}
\end{table}

Stream parameters that survive plain video re-encoding tell a similar story.
Delivery-parameter matching (DPM; Sec.~\ref{sec:audit_screens}) inspects a nine-field tuple covering width, height, nominal frame rate, packet-counted frames, audio presence, codec, sample rate, channel count, and audio-minus-video duration $d$ (Tab.~\ref{tab:tuples}). %
Every \emph{Main} endpoint emits deterministic video geometry, frame rate, and frame count across all of its deliveries, and every reference clip in each 720p DPM training fold shares one identical nine-field tuple (48/48 per fold for Seedance~2.5 and Grok~1.5; 32/32 for Wan~3.0~Prime and Kling~3.0~Std). %
While pinned-policy C2PA authenticates all 160 \emph{Main} clips from the three Seedance~2.x tiers (187 clips including \emph{Low-res}), the remaining seven \emph{Main} pools rely on archived gateway-labelled deliveries. %
As Tab.~\ref{tab:tuples} illustrates, distinct pipelines can differ across every field (Kling~3.0~Std returns $1920{\times}1080$ on 720p requests) or by a single one-frame muxing offset $d$ (Grok~1.5 vs.\ 1.5~Lite), whereas shared-pipeline 720p tiers emit identical tuples and Seedance~2.5 vs.\ 2.0~Fast diverge in geometry only at 480p ($854{\times}480$ vs.\ $864{\times}496$, 97 silent frames; \prelim{preliminary at $n = 9$}). %
Because a provider can rewrite stream tuples without changing the underlying model, DPM screens only unnormalised packaging defaults (Sec.~\ref{sec:audit_screens}). %

\subsection{Processing-Dependent, Pair-Specific Visual Separability}\label{sec:visual_content}

To measure what decoded pixels still reveal once packaging is normalised across all 45 \emph{Main} pairs (40 or 60 shared prompts per pair), we train logistic-regression (LR) probes on three feature families within each processing condition, making no claim of cross-condition transfer. %
The \emph{semantic} probe mean-pools SigLIP~\citep{siglip2023} frame embeddings (capturing rendered visual style as well as scene content), and the \emph{video} probe uses X-CLIP~\citep{xclip2022} embeddings evaluated at N2 and the 480p CRF-32 round trip. %
Over 16 evenly spaced frames, the 110-dimensional \emph{low-level} probe extracts a 64-bin radial log power spectrum and 20 block-DCT statistics (means and variances of 10 low-order AC magnitudes) from the $3\times 3$ luma median-filter residual, concatenated with 4 first- and second-order temporal luma-difference statistics and 22 colour statistics (RGB means/standard deviations and a 16-bin saturation histogram), so it spans frequency, temporal, and colour cues rather than pure sensor-style noise. %
Features are standardised inside each training fold, with regularisation chosen from $\{0.01, 0.1, 1, 10\}$ via inner three-fold prompt-grouped log-loss search, evaluated under outer five-fold prompt-grouped cross-validation, and paired with 2,000-resample prompt-cluster bootstrap confidence intervals. %

We compare pairwise separability before and after uniform re-encoding (Fig.~\ref{fig:visual_findings}b, defining ``format-identical'' on the four video fields: width, height, nominal frame rate, and frame count). %
Across all 45 pairs, the median shift $\text{AUROC}(\text{N1}) - \text{AUROC}(\text{N2})$ is modest: 0.008 for semantic features and 0.014 for low-level features (0.007 across the 22 format-identical pairs vs.\ 0.028 across the 23 format-distinct pairs). %
The largest low-level losses were concentrated in pairs involving native-1080p deliveries. In a post-hoc analysis, the nine $1920{\times}1080$ Kling~3.0~Std pairs fall from 0.889--0.984 low-level AUROC at N1 to 0.712--0.934 at N2 (median drop 0.188), whereas their semantic AUROC barely moves (median shift $-0.014$) and the 14 non-Kling pairs involving 30-fps Wan shift by a median of only 0.014 on low-level features. %

\begin{table}[t]
    \centering
    \caption{\textbf{Pairwise AUROC at N2 across the four disjoint pair categories in \emph{Main}} (45 pairs total; median [min, max]).} %
    \label{tab:pair_categories}
    \footnotesize
    \setlength{\tabcolsep}{2.5pt}
    \adjustbox{max width=\linewidth}{%
\begin{tabular}{@{}lrccc@{}}
        \toprule
        Pair category & Pairs & Semantic & Low-level & Video \\
        \midrule
        Seedance cross-version & 5 & 0.966 [0.789, 0.994] & 0.918 [0.749, 0.997] & 0.707 [0.619, 0.762] \\ %
        Cross-vendor & 35 & 0.929 [0.771, 0.986] & 0.852 [0.637, 0.999] & 0.719 [0.610, 0.808] \\ %
        Alibaba cross-family & 2 & 0.810 [0.804, 0.816] & 0.785 [0.757, 0.813] & 0.704 [0.660, 0.749] \\ %
        Same-version tiers & 3 & 0.770 [0.508, 0.825] & 0.781 [0.551, 0.863] & 0.611 [0.548, 0.649] \\ %
        \bottomrule
    \end{tabular}%
}
\end{table}

At N2, separability follows a clear hierarchy across the four pair categories in Tab.~\ref{tab:pair_categories} (Fig.~\ref{fig:visual_findings}a): probes separate Seedance cross-version pairs and cross-vendor pairs cleanly (semantic medians 0.966 and 0.929), drop to intermediate AUROC between Alibaba's HappyHorse and Wan families, and are weakest on the three same-version tier pairs (medians 0.770 semantic and 0.781 low-level), while X-CLIP trails in every group (medians $\le 0.719$). %
Category ranges nonetheless overlap: cross-version Seedance~2.5 vs.\ 2.0~Fast (0.789 semantic, 0.749 low-level, well below 0.926 and 0.943 on 2.5 vs.\ 2.0~Mini) is less separable than same-version Grok~1.5 vs.\ 1.5~Lite (0.825 and 0.781). %
In a post-hoc tally across all 45 pairs, semantic embeddings outperform low-level features on 37/45 pairs at N2, leaving 8/45 low-level leads (mostly involving Seedance~2.0~Mini and none involving Kling), down from 14/45 low-level leads at N1 before Kling's native resolution was normalised. %

\begin{table}[t]
    \centering
    \caption{\textbf{Same-version tier separability at N2} (720p, 24\,fps, CRF~23; prompt-grouped 5-fold CV). Cells report out-of-fold AUROC and positive clips detected at the empirical ROC operating point with FPR at most 1\%; brackets give the AUROC 95\% confidence interval. The first listed model is class~0 and the second is the positive class; with 40 or 60 class-0 clips, the reported operating points have zero observed false positives (0/40 or 0/60), which is not a certified deployment false-positive rate. Video probe (X-CLIP) AUROC is 0.548, 0.611, and 0.649.} %
    \label{tab:tiers}
    \footnotesize
    \setlength{\tabcolsep}{3pt}
    \adjustbox{max width=\linewidth}{%
\begin{tabular}{@{}lcc@{}}
        \toprule
        Tier pair (clips; labels) & Semantic (SigLIP) & Low-level (110-d LR) \\ %
        \midrule
        Wan~3.0~Prime / 3.0 & 0.508; 0/40 & 0.551; 1/40 \\ %
        \quad (40+40; Gateway) & [0.436, 0.581] & [0.471, 0.632] \\ %
        \addlinespace[1.5pt]
        Seedance~2.0~Mini / 2.0~Fast & 0.770; 1/40 & 0.863; 9/40 \\ %
        \quad (40+40; Auth.\ C2PA) & [0.685, 0.859] & [0.782, 0.933] \\ %
        \addlinespace[1.5pt]
        Grok~1.5 / 1.5~Lite & 0.825; 30/60 & 0.781; 2/60 \\ %
        \quad (60+60; Gateway) & [0.767, 0.884] & [0.708, 0.849] \\ %
        \bottomrule
    \end{tabular}%
}
\end{table}

Even among the three same-version tier pairs, different cues carry the signal (Tab.~\ref{tab:tiers}). %
Grok~1.5 vs.\ 1.5~Lite separates primarily in semantic appearance (0.825 AUROC, catching 30/60 Lite clips at 0/60 observed false positives), whereas C2PA-authenticated Seedance~2.0~Mini vs.\ 2.0~Fast separates more strongly in low-level statistics (0.863 AUROC and 9/40 Fast clips caught at 0/40 false positives, vs.\ 0.770 semantic). %
Wan~3.0~Prime vs.\ 3.0, by contrast, remains near chance ($\le 0.551$ at both N1 and N2, and $\le 0.591$ even under optimistic selection across all 17 evaluated attribution configurations). %
Heavier compression further erodes these signals: retraining within the 480p CRF-32 round trip lowers semantic median AUROC to 0.899 on cross-vendor pairs and 0.667 on same-version tiers (with Seedance~2.0~Mini / 2.0~Fast dropping from 0.770 to 0.667), while Seedance~2.5 / 2.0~Mini low-level AUROC drops from 0.943 to 0.826 and low-FPR detections fall from 29/60 to 3/60. %

Multi-class and open-set attribution face similar ceilings.
Across eight feature families and 17 configurations on the six-model, 360-clip \emph{Main} subset at N2, six-way macro-F1 peaks at 0.613 ($[0.575, 0.654]$) with SigLIP mid-frame LR (\prelim{our reproduction of SAGA~\citep{saga2026} reached 0.871 accuracy on its source benchmark vs.\ a 0.95 gate, yielding 0.453 six-way macro-F1 in its paper configuration and 0.538 tuned on \emph{Main}, and remains preliminary}). %
When that N2-trained SigLIP classifier is evaluated on degraded clips without retraining, macro-F1 drops by 0.058 at CRF~28 and 0.188 after the 480p CRF-32 round trip, while maximum-softmax open-set rejection of unseen substitutes reaches at most 0.623 AUROC using DINOv3~\citep{dinov3_2025} mean embeddings. %

Finally, label authenticity limits how passive separability on unsigned APIs can be interpreted. %
Only the three multi-clip Seedance~2.x tiers in \emph{Main} (160/160 clips, or 187/187 with \emph{Low-res}; Sec.~\ref{sec:provenance}) carry tier-authenticated signatures under our pinned policy. %
Elsewhere, reference labels are unverified gateway claims, meaning near-chance separability on Wan~3.0~Prime vs.\ 3.0 ($\le 0.591$) is equally consistent with minimal visual difference between tiers, a shared upstream generator, or substitution during data collection. %

\section{What Deployed Provenance Establishes}\label{sec:provenance}

\subsection{Four Authentication Levels and Pinned-Policy Verification}\label{sec:prov_levels}
Interpreting cryptographic provenance~\citep{c2paspec,c2pafall2026} in an API setting requires separating four levels of assurance:
\begin{itemize}
    \item \textbf{Level~1 (Cryptographic validity):} parsing the manifest store and checking both claim-signature validity (\texttt{claimSignature.validated}) and \texttt{bmffHash} payload binding, which establish byte integrity since signing but do not rule out self-signed per-clip CAs or untrusted third-party signers. %
    \item \textbf{Level~2 (Signer trust):} the verifier checks the certificate chain and leaf organisation against the policy for that vendor. Each policy specifies its trust anchors and reference time. The BytePlus policy also includes a separate leaf revocation check. These checks establish the signer's organisational identity under the buyer's policy. %
    \item \textbf{Level~3 (Claim consistency):} matching the signed identity assertion (such as \texttt{model\_name}) to the ordered endpoint at the signed granularity and checking the active manifest's instance ID against prior authenticated billed jobs, establishing whether the vendor's claim matches the order, contradicts it, or duplicates an earlier delivery. %
    \item \textbf{Level~4 (Internal computation):} model weights, denoising steps, internal resolution, and quantisation, which no file-carried signature can attest. %
\end{itemize}

Our verifier $F$ applies Levels~1--3 to raw delivered bytes (Appendix~\ref{app:verifier}). It checks the claim signature and \texttt{bmffHash} binding, validates the signer and certificate dates under the vendor-specific policy, compares the signed claim with the order at the granularity it supports, and checks the instance ID against prior authenticated billed jobs. %

Applying $F$ to all 187 audited BytePlus deliveries across Seedance~2.5, 2.0~Fast, and 2.0~Mini in \emph{Main} and \emph{Low-res} verified 187/187 files with signed \texttt{model\_name} matching the ordered tier (alongside four verified \emph{Census} clips, including \prelim{Seedance~2.0 at $n = 1$}, for 191 verified files in total). %
Every BytePlus certificate chain terminates at \emph{GlobalSign Secure Mail Root R45} via \emph{Byteplus Pte.\ Ltd.}, a root included in the Mozilla store (e-mail trust bit) rather than the C2PA trust list, which makes signer trust an explicit buyer policy decision rather than an automatic ecosystem guarantee. %
The inspected Seedance manifests also record a signer-asserted action time (\texttt{when}) rather than a trusted RFC~3161 timestamp; $F$ therefore evaluates certificate validity at \texttt{when} and runs a separate audit-time OCSP check on the leaf (\texttt{GOOD}, signature-verified), though neither proves when signing actually occurred. %
Across the audited population, $F$ observed 0/187 duplicate instance IDs (all 6/6 seeded replays carried fresh IDs) and classified all 18/18 adversarial controls (\texttt{mdat} byte flips, cross-tier manifest transplants, three forged signer chains, raw cheaper-tier swaps, CRF-23 re-encodes, and re-presented originals) into their expected rejection classes and failure codes (Appendix~\ref{app:verifier}). %

\subsection{Cross-Regime Census, Fragility, and Missing Bindings}\label{sec:prov_census}

\begin{table*}[t]
    \centering
    \caption{\textbf{Provenance census across 19 commercial video models (605 deliveries: 335 C2PA, 209 GB~45438, 61 none; two byte-identical re-extractions).} Each model used one consistent regime across \emph{Main}, \emph{Census}, and \emph{Low-res}; \prelim{the 9 single-clip \emph{Census} models ($n = 1$) and second-session re-checks ($n = 2$ sessions)} are preliminary.} %
    \label{tab:census}
    \footnotesize
    \setlength{\tabcolsep}{3pt}
    \begin{tabular}{@{}>{\raggedright\arraybackslash}p{0.21\textwidth}l>{\raggedright\arraybackslash}p{0.17\textwidth}l>{\raggedright\arraybackslash}p{0.39\textwidth}@{}}
        \toprule
        Vendor: models & Regime & Identity field(s) & Granul. & Third-party verifiability \\
        \midrule
        BytePlus: Seedance 2.5, 2.0~Fast, 2.0~Mini; \prelim{2.0 ($n=1$)} & C2PA & \texttt{model\_name}, \texttt{when}, ID & Tier & Verified 187/187 across 3 tiers (tier = order; \prelim{+1/1 on 2.0, $n=1$}); GlobalSign R45 (Mozilla e-mail, not C2PA list); OCSP \texttt{GOOD} \\ %
        \prelim{Black Forest Labs: FLUX.3 video ($n=1$)} & \prelim{C2PA} & \prelim{\texttt{softwareAgent} FLUX.3, ID} & \prelim{Model} & \prelim{Pinned-policy verified at model granularity; chains to GlobalSign R45} \\ %
        \prelim{Runway: Gen-4.5 ($n=1$)} & \prelim{C2PA} & \prelim{agent + version 4.5.0} & \prelim{Agent} & \prelim{Not verified: \texttt{signingCredential.invalid} despite C2PA-listed root} \\ %
        xAI: Grok Imagine 1.5, 1.5~Lite & C2PA & shared ``Grok Imagine'' & Product & Not authenticated under our pinned policy: 142/142 distinct per-clip self-signed CAs (\texttt{LOCAL USE ONLY}) \\ %
        Alibaba: \prelim{Wan 2.6, 2.7 ($n=1$ each)}, 3.0, P; HH~1.1 & GB 45438 & shared producer, \texttt{ProduceID} & Vendor & Structure 164/164; buyer integrity untested (scheme/key not supplied); prefix matches 24/24 own \& other tier \\ %
        Kuaishou: Kling \prelim{3.0~Pro ($n=1$)}, Std & GB 45438 & vendor ID, \texttt{ProduceID}, \prelim{\texttt{sm3} (prelim.)} & Vendor & Structure 42/42; buyer integrity untested (scheme/key not supplied); \prelim{2nd session added self-signed C2PA ($n=1$)} \\ %
        \prelim{MiniMax: Hailuo 2.3, 3, 3-max ($n=1$ each)} & \prelim{GB 45438} & \prelim{vendor ID, \texttt{ProduceID}, \texttt{DS}} & \prelim{Vendor} & \prelim{Structure 3/3}; buyer integrity untested (scheme/key not supplied) \\ %
        ByteDance: Seedance 1.5~Pro & None & n/a & None & None (61/61 files) \\ %
        \bottomrule
    \end{tabular}
\end{table*}

Stepping across the full 19-model census in Tab.~\ref{tab:census} (605 deliveries, including \prelim{nine single-clip models at $n = 1$}) highlights four practical deployment limits. %

\paragraph{Metadata presence rarely yields tier authentication.}
Although 18 of the 19 endpoints attach metadata (8 C2PA, 10 GB~45438, and 1 with none), only BytePlus signs verifiable tier claims under our pinned policy. %
Among the other C2PA deployments, Grok signs the shared product label ``Grok Imagine'' on both tiers using 142/142 distinct per-clip self-signed Ed25519 CAs marked \texttt{LOCAL USE ONLY}; \prelim{FLUX.3 verifies under our pinned policy only at model granularity ($n = 1$); and Runway Gen-4.5 signs software-agent version 4.5.0 (whose equivalence to model tier is unestablished) while failing validation with \texttt{signingCredential.invalid} despite chaining to a C2PA-listed root ($n = 1$)}. %

\paragraph{GB~45438-2025 tags remain vendor-level and informational.}
All 209 deliveries across the 10 GB~45438-2025~\citep{gb45438,cacmeasures2025} endpoints parse cleanly across all seven visible JSON keys, with a non-empty \texttt{ProduceID} unique to each billed job within its vendor (209/209). %
For an API buyer, however, each vendor reuses a single \texttt{ContentProducer} string across models and tiers, vendor integrity schemes and external verification keys were not available for buyer-side testing, and Alibaba's 66-character \texttt{ReservedCode1} prefix is identical across all 24 Wan~3.0 and 24 Wan~3.0~Prime clips inspected. %

\paragraph{Credentials are fragile under routine transcoding.}
\prelim{In our preliminary cross-vendor survival matrix ($n \le 3$ clips per vendor and transform; Appendix~\ref{app:verifier}), FFmpeg remuxing, metadata stripping, and CRF-23 or CRF-28 re-encoding stripped C2PA manifests completely, leaving 0/3 Seedance and 0/1 FLUX.3 clips verifiable even when explicit metadata-copy flags were supplied.} %
\prelim{GB~45438 user-data tags likewise disappeared by default under remuxing and re-encoding (0/3 per vendor), surviving only when explicit metadata-tag copying was enabled (3/3 per vendor).} %

\paragraph{Manifests omit request binding and can shift over time.}
Because the inspected Seedance manifest store embeds no prompt, request parameters, or client nonce, a valid signature alone cannot detect cross-buyer replay of a genuine clip. %
\prelim{Moreover, in our preliminary second session eight days later ($n = 2$ sessions, one clip per model across six \emph{Census} endpoints), 5/6 re-tested models retained their original regime and identity fields, while Kling~3.0~Pro newly attached a self-signed vendor-level C2PA manifest (``Kling AI'') alongside GB~45438.} %

%% file: sections/05_experiments.tex
\section{Graded Auditing Across the Gap}\label{sec:audit}

In the audit evaluation that follows, a \emph{signed endpoint} refers to an ordered endpoint whose tier claims can be authenticated under our pinned policy (specifically, the three multi-clip Seedance~2.x tiers, plus \prelim{Seedance~2.0 at $n = 1$}). %
\emph{Unsigned} refers to all endpoints lacking tier-authenticated credentials under that policy, including Grok's self-signed C2PA and the GB~45438 endpoints. %

\subsection{Why Passive Classifiers Stay Outside the Audit Chain}\label{sec:audit_rules}
Suppose an audit inspects $n = 30$ independent deliveries from an endpoint where an intermediary substitutes a fraction $p$ of clips. %
If a candidate per-clip screen catches a substitute with probability $d$ and falsely flags an honest clip with probability $q$, each delivery triggers with probability $\pi = pd + (1 - p)q$. %
When an audit escalates at $\ge k$ triggers out of $n$, both its escalation rate and the zero-failure honest calibration sample $N(q)$ needed to certify $q$ under a one-sided 95\% Clopper--Pearson upper bound~\citep{clopper1934} follow directly: %
\begin{equation}\label{eq:power}
\begin{aligned}
    P(\Bin(n, \pi) \ge k) &= \sum_{j=k}^{n} \binom{n}{j} \pi^j (1 - \pi)^{n-j}, \\ %
    N(q) &= \left\lceil \frac{\ln 0.05}{\ln(1 - q)} \right\rceil. %
\end{aligned}
\end{equation}

Under our partial-mixing benchmark ($n = 30$, $p = 10\%$, 95\% recall, and at most 1\% endpoint false accusation), requiring $k = 2$ triggers caps recall at 81.6\% even for an oracle detector with $d = 1$ and $q = 0$ (reaching 95\% recall at $k = 2$ would require $n = 46$). %
Achieving 95\% recall within $n = 30$ clips therefore forces a single-clip trigger ($k = 1$), which demands $q_1 \le 0.0335\%$ and $N(q_1) = 8{,}943$ zero-failure honest calibration clips per endpoint (\$1.17k--\$10.4k at pilot clip prices, or 299 clean 30-clip audits totalling 8,970 clips). %

Against \emph{wholesale} substitution ($p = 1$), calibration sample size is much smaller in principle: a 6-of-30 trigger rule at $q = 5\%$ yields a 0.33\% false-escalation rate, needs only $N(0.05) = 59$ zero-failure honest clips, and achieves 99.98\% recall even at $d = 0.5$. %
Yet even this wholesale test presupposes three conditions that are not established for the intended deployment setting. %
It requires an authentic honest calibration pool, whereas unsigned reference clips carry only unverified gateway labels; it assumes a known substitute class, whereas maximum-softmax open-set rejection of unseen substitutes reaches at best 0.623 AUROC; and it assumes cross-session stationarity, which our measurements do not establish (failed seeded replay cannot separate seed non-determinism from temporal drift). %
We therefore keep passive visual classifiers out of audit triggers altogether, admitting ARM and DPM solely as review-grade screens because they require no model of the substitute and produced zero false flags on our honest pools. %

\begin{table*}[t]
    \centering
    \caption{\textbf{Verdict lattice and buyer actions in the provenance-first graded audit.} Verdicts follow strict priority $\mathrm{AC} > \mathrm{AD} > \mathrm{BR} > \mathrm{RV} > \mathrm{CC} > \mathrm{AB}$; an original-file demand is a pending contractual action under a pass-through clause whose non-cure yields $\mathrm{BR}$.} %
    \label{tab:verdicts}
    \footnotesize
    \setlength{\tabcolsep}{3pt}
    \adjustbox{max width=\linewidth}{%
\begin{tabular}{@{}ll>{\raggedright\arraybackslash}p{0.31\textwidth}l>{\raggedright\arraybackslash}p{0.31\textwidth}@{}}
        \toprule
        Code & Name & Triggering evidence and contract requirement & Grade & Buyer next action (evaluated case) \\
        \midrule
        $\mathrm{AC}$ & Auth.\ contradiction & $F$ verifies pinned-policy signature; known signed tier differs from order ($k = 1$) & Accuse & Dispute job with signed file (100/100 raw 720p tier swaps; \prelim{9/9 at 480p}) \\ %
        $\mathrm{AD}$ & Auth.\ duplicate & $F$ verifies pinned-policy signature; signed instance ID repeats across separately billed jobs under a fresh-generation contract, excluding retries ($k = 1$) & Accuse & Dispute duplicate billing (10/10 raw signed archive reuses) \\ %
        Demand $\to \mathrm{BR}$ & Original demand $\to \text{Contract breach}$ & Signed-endpoint delivery is unverifiable under $F$ with an agreed pass-through clause; proposed $\mathrm{BR}$ if demand goes uncured (untested) or $\ge 4/30$ clips are unverifiable & Demand / Breach & Demand ordered tier's vendor original linked via ARM; declare breach if uncured or $\ge 4/30$ (1,020/1,020 720p proc./cross-vendor cases; \prelim{72/72 at 480p}) \\ %
        $\mathrm{RV}$ & Review & ARM archive match or DPM stream-tuple mismatch; flags never combine into an accusation & Review screen & Sampled human inspection (580/580 unnorm.\ unsigned tuple mismatches; 0/100 in the separate normalised unsigned test; 130/260 unsigned reuses) \\ %
        $\mathrm{CC}$ & Claim-consistent & $F$ verifies pinned-policy signature; signed tier matches order and instance ID is fresh & Claim verified & Accept delivery; establishes vendor claim, not internal compute (187/187 genuine Seedance clips) \\ %
        $\mathrm{AB}$ & Abstain & Unsigned endpoint with neither ARM nor DPM triggered (GB~45438 logged as information only) & Abstain & Log metadata; draw no identity conclusion (280/280 shared-tuple Wan cases) \\ %
        \bottomrule
    \end{tabular}%
}
\end{table*}

To audit an endpoint, the buyer archives raw delivered bytes alongside request and billing ledgers prior to any local transcoding, runs the pinned-policy verifier $F$ (Sec.~\ref{sec:prov_levels}), and assigns the highest-priority verdict in $\mathrm{AC} > \mathrm{AD} > \mathrm{BR} > \mathrm{RV} > \mathrm{CC} > \mathrm{AB}$ (Tab.~\ref{tab:verdicts}). %
Only $\mathrm{AC}$ (authenticated tier contradiction) and $\mathrm{AD}$ (authenticated instance-ID duplicate across separately billed fresh jobs, excluding retries and idempotent returns) reach accusation grade at $k = 1$, while order-matching signatures return $\mathrm{CC}$, confirming claim agreement rather than internal computation. %
Under a signed-endpoint pass-through clause, any unverifiable delivery triggers a pending demand for the ordered tier's signed vendor original linked to the delivery via ARM; geometric or temporal edits break that ARM link, though edited deliveries already violate pass-through. %
We propose escalating to contract breach ($\mathrm{BR}$) if a demand goes uncured (cure was not evaluated) or if at least 4 of 30 deliveries arrive unverifiable. %
Elsewhere, ARM and DPM emit bounded review flags ($\mathrm{RV}$) that never accumulate into an accusation, GB~45438 tags are logged for information only, and untriggered unsigned clips abstain ($\mathrm{AB}$). %
Over $M$ deliveries, false accusations are bounded by $P(\text{false accusation}) \le \varepsilon_{\text{trust}} + M(\varepsilon_F + \varepsilon_D)$, capturing unquantified trust-policy, verifier, and order-mapping errors without any statistical classifier term. %

\subsection{Review-Grade Screens: Archive Reuse and Delivery Defaults}\label{sec:audit_screens}
\paragraph{Archive reuse matching (ARM).}
ARM checks for cached clip reuse (S5) against the buyer's own prior deliveries, grounding its reference pool in physical possession rather than gateway labels. %
We build a video hash from image-hashing primitives~\citep{phash2010}, combining 63 spatial bits from the middle sampled frame with $64m$ thumbnail bits from $m \le 17$ sampled frames. We pair it with 8-fps, 64-px-wide SSIM~\citep{ssim2004}; a match requires normalised Hamming distance $\le 0.05$, SSIM $\ge 0.90$, and equal hash length (pairs with unequal $m$ cannot match, are marked non-comparable in negative checks while remaining in total pair counts, and count as misses on planted copies; Appendix~\ref{app:arm}). %
Across all 586 originals in \emph{Main} and \emph{Low-res} (171,405 total pairs), ARM produced zero false matches across 133,359 comparable cross-prompt pairs (of 168,081), all 48 same-model regenerations, and 2,245 comparable other same-prompt pairs (of 3,276). %
Evaluated on the 390 planted archive-reuse cases in \emph{Substitution} (10 signed Seedance and 20 unsigned sources across 13 transforms), $F$ abstains on all 260 unsigned reuses while ARM flags 130/260 for review: it catches almost all unedited, CRF-23/28, 480p-downscaled, and colour-shifted copies (116/120) and over half at 5\% speed-up (12/20), but is largely bypassed by spatial crops, temporal trims, 10\% speed-ups, and horizontal flips (learned copy detectors such as ViSiL~\citep{visil2019} were not evaluated; per-transform counts in Tab.~\ref{tab:arm_perturbations}). %
On the 130 signed-endpoint reuses, $F$ directly proves all 10/10 raw duplicates ($\mathrm{AD}$) and triggers original-file demands on all 120/120 processed copies, where ARM identifies 57/120 as archived reuses without adding escalations. %

\paragraph{Delivery-parameter matching (DPM).}
DPM compares the nine stream parameters of a delivery with the most frequent tuple(s) in a prompt-disjoint reference set for the claimed endpoint (Tab.~\ref{tab:tuples}). At 720p the reference excludes the delivery's entire prompt fold. At 480p it excludes every clip from the same base prompt. Eight fields must match exactly, and the audio-minus-video offset $d$ must agree within $500\,\mu\text{s}$ (Appendix~\ref{app:arm}). %
In our evaluated substitution pairs, the signed-endpoint DPM reference pools consist of 60 C2PA-authenticated \emph{Main} Seedance~2.5 clips and \prelim{nine \emph{Low-res} Seedance~2.5 clips ($n = 9$)}, whereas the Grok~1.5, Wan~3.0~Prime, and Kling~3.0~Std reference pools use archived gateway-labelled clips. %
DPM raised zero false flags on the honest pools in \emph{Substitution} (0/200 at 720p and \prelim{0/9 at 480p, $n = 9$}). %
\prelim{DPM raised no flags in 27 format-matched comparisons across collection times. Seven compared \emph{Main} with one \emph{Census} clip per endpoint (three audio mismatches excluded). Eighteen compared requests repeated about 3~h apart on six \emph{Low-res} endpoints, with three prompts each (Seedance~2.0~Mini uncollected). Two compared one clip per session on FLUX.3 and Wan~2.6 eight days apart (four duration or resolution mismatches excluded).} %
At the same time, DPM cannot separate shared-tuple tiers and is easily bypassed by adaptive remuxing or transcoding: a single normalisation pass reduces DPM detections from 160/160 to 0/160 across the three 720p tuple-distinct pairs with zero honest false flags (\prelim{and 480p cropping reduces 480p Seedance~2.5 $\leftarrow$ Fast flags from 9/9 to 0/9, $n = 9$}). %
Dropping or trimming audio on HappyHorse and Grok~1.5~Lite takes $\le 0.44$ median CPU-s while leaving decoded video bit-identical (\texttt{framemd5}), and transcoding Wan~3.0 to Kling~3.0~Std's tuple takes 16.3 median CPU-s at 47.6~dB mean PSNR (0.993 SSIM), confirming that DPM belongs strictly at review grade ($\mathrm{RV}$). %

\subsection{Ground-Truth Substitution Evaluation}\label{sec:audit_eval}

\begin{table*}[t]
    \centering
    \caption{\textbf{Per-clip ground-truth evaluation across \emph{Substitution}, one-pass normalisation, and archive reuse.} Honest pools produced 0 false flags across $F$ (187/187 verified), DPM (0/200 at 720p; \prelim{0/9 at 480p, $n = 9$}), and ARM: ARM had zero matches among 133,359 comparable cross-prompt pairs, 48 same-model regenerations, and 2,245 comparable other same-prompt pairs drawn from 586 clips; total pair counts and comparability exclusions are reported in the text. Rows report individual checks on overlapping slices and must not be summed across blocks. ``Signed'' describes the claimed endpoint under our pinned policy; a DPM review flag ($\mathrm{RV}$) never supersedes an original-file demand or an accusation-grade $F$ verdict.} %
    \label{tab:d4_outcomes}
    \footnotesize
    \setlength{\tabcolsep}{3pt}
    \adjustbox{max width=\linewidth}{%
\begin{tabular}{@{}>{\raggedright\arraybackslash}p{0.28\textwidth}lrrcc@{}}
        \toprule
        Slice / claimed $\leftarrow$ substitute pair & Endpoint status & Src.\ & Variants (cases) & Provenance $F$ & DPM / ARM screen \\
        \midrule
        \multicolumn{6}{@{}l}{\emph{Fresh-substitution matrix (2,061 cases, 6 model pairs, 7 pair--resolution settings; 0/1,201 signed-endpoint cases claim-consistent)}} \\ %
        Seedance~2.5 $\leftarrow$ 2.0~Mini & Signed & 60 & 1 raw + 6 proc.\ (420) & 60/60 $\mathrm{AC}$; 360/360 Demand & DPM: 0/420 (shared tuple) \\ %
        Seedance~2.5 $\leftarrow$ 2.0~Fast & Signed & 40 & 1 raw + 6 proc.\ (280) & 40/40 $\mathrm{AC}$; 240/240 Demand & DPM: 0/280 (shared tuple) \\ %
        Seedance~2.5 $\leftarrow$ HappyHorse~1.1 & Signed & 60 & 7 raw/proc.\ (420) & 420/420 Demand & DPM: 420/420 $\mathrm{RV}$ (unnorm.) \\ %
        \quad \emph{720p signed subtotals} & \emph{Signed} & --- & \emph{100 raw; 1,020 proc./cross} & \emph{100/100 $\mathrm{AC}$; 1,020/1,020 Demand} & \emph{DPM: 420/420 on cross-vendor} \\ %
        Grok~1.5 $\leftarrow$ 1.5~Lite & Unsigned & 60 & 5 raw/proc.\ (300) & 0/300 & DPM: 300/300 $\mathrm{RV}$ (unnorm.) \\ %
        Kling~3.0~Std $\leftarrow$ Wan~3.0 & Unsigned & 40 & 7 raw/proc.\ (280) & 0/280 & DPM: 280/280 $\mathrm{RV}$ (unnorm.) \\ %
        \quad \emph{Unsigned diff.-tuple subtotal} & \emph{Unsigned} & --- & \emph{12 variants (580)} & \emph{0/580} & \emph{DPM: 580/580 $\mathrm{RV}$ (unnorm.)} \\ %
        Wan~3.0~Prime $\leftarrow$ Wan~3.0 & Unsigned & 40 & 7 raw/proc.\ (280) & 0/280 & DPM: 0/280 (shared tuple) \\ %
        \addlinespace[1.5pt]
        \prelim{480p: Seedance~2.5 $\leftarrow$ 2.0~Fast ($n=9$)} & \prelim{Signed (prelim.)} & \prelim{9} & \prelim{1 raw + 8 proc.\ (81)} & \prelim{9/9 $\mathrm{AC}$; 72/72 Demand} & \prelim{DPM: 63/63 uncrop.; 0/18 crop.} \\ %
        \midrule
        \multicolumn{6}{@{}l}{\emph{Adaptive one-pass parameter normalisation (reuses source clips of the three 720p tuple-distinct pairs above)}} \\ %
        Norm.\ signed cross-vendor (HH$\to$2.5) & Signed & 60 & 1-pass audio drop (60) & 60/60 Demand & DPM: 0/60 (was 60/60) \\ %
        All three normalised pairs (160; includes the 60 above) & Signed + Unsigned & 160 & 1-pass norm.\ (160) & Demand on 60; 0/100 unsig. & DPM: 0/160 (was 160/160) \\ %
        \midrule
        \multicolumn{6}{@{}l}{\emph{Archive-reuse evaluation across 13 transforms (reuses 10 signed Seedance and 20 unsigned source clips)}} \\ %
        Unsigned archive reuse & Unsigned & 20 & 13 raw/proc.\ (260) & 0/260 & ARM: 130/260 $\mathrm{RV}$ \\ %
        Signed archive reuse, raw & Signed & 10 & 1 raw (10) & 10/10 $\mathrm{AD}$ & ARM: 10/10 (superseded by $\mathrm{AD}$) \\ %
        Signed archive reuse, processed & Signed & 10 & 12 proc.\ (120) & 120/120 Demand & ARM: 57/120 reuse ID (0 extra esc.) \\ %
        \bottomrule
    \end{tabular}%
}
\end{table*}

Table~\ref{tab:d4_outcomes} brings together per-clip outcomes across all 2,061 fresh-substitution cases in \emph{Substitution} (spanning unmodified delivery, remuxing, metadata stripping, CRF-23/28 re-encoding, C2PA or GB~45438 transplant, and \prelim{preliminary 480p cropping at $n = 9$}), alongside the 160-clip normalisation test and 390-clip archive-reuse test. %
Not one of the 1,201 signed-endpoint substitutes passed verification as claim-consistent (0/1,201): every unmodified tier swap produced an accusation-grade contradiction ($\mathrm{AC}$), while every post-processed or cross-vendor substitute failed verification and triggered an original-file demand. %
On unsigned endpoints, DPM flagged all 580/580 unnormalised tuple-distinct substitutes for review, whereas a single normalisation pass reduced DPM flags from 100/100 to 0/100 on the unsigned Grok and Kling pairs and from 60/60 to 0/60 on the signed-endpoint HappyHorse-to-Seedance pair (which remains caught by $F$'s original-file demand); shared-tuple Wan~3.0~Prime $\leftarrow$ Wan~3.0 triggered neither check (0/280). %

At the 30-clip audit level across all 40 multi-prompt 720p fresh-substitution scenarios (Appendix~\ref{app:audit_simulation}), deterministic per-clip detection makes audit-level behaviour follow Eq.~\eqref{eq:power} directly. %
At a $p = 10\%$ mixing rate, the analytic probability of raising review-or-higher (Review+, covering $\mathrm{AC}$, $\mathrm{AD}$, $\mathrm{BR}$, pending demands, and $\mathrm{RV}$) is 95.8\% across the 33 detecting scenarios, and the automatic-breach rate ($\ge 4/30$ unverifiable clips) is 35.3\% across the 19 processed or cross-vendor signed-endpoint scenarios; 10,000 Monte Carlo audits per scenario closely match both figures (0.9563--0.9587 and 0.3476--0.3524), with 0/10,000 false flags in every honest ($p = 0$) scenario. %
Adding ARM left fresh-substitution verdicts unchanged across all 40 scenarios, contributing its 130/260 review flags specifically against unsigned archive reuse. %

%% file: sections/06_limitations.tex
\section{Limitations}\label{sec:limitations}

\paragraph{Scope.}
All deliveries in \emph{Main}, \emph{Census}, and \emph{Low-res} were collected through a single gateway (OpenRouter) without direct vendor API contracts, primarily in one session (\prelim{plus a preliminary six-model re-check eight days later at one clip per model, $n = 2$ sessions}). %
Our measurements therefore describe the tested endpoints at collection time rather than the broader commercial video industry.
Only BytePlus signed verifiable tiers across multi-clip Seedance~2.x endpoints (\prelim{plus Seedance~2.0 at $n = 1$}), and \prelim{provenance configurations can evolve across sessions, as illustrated by Kling~3.0~Pro adding a self-signed C2PA manifest ($n = 1$) in the re-check}. %

\paragraph{Evidence grade.}
On endpoints without a tier-authenticating credential under our pinned policy, the buyer cannot independently verify the ordered tier. A credential may still authenticate a vendor's model-level claim, as in the single FLUX.3 delivery. Passive labels and unsigned DPM references rely on gateway claims. DPM checks stream defaults, and ARM uses fixed-grid perceptual matching. %
Similarly, GB~45438-2025 tags~\citep{gb45438} are designed for regulatory traceability rather than buyer-side tier verification; their integrity schemes and external verification keys were not supplied (leaving buyer verification untested), and Wan~3.0 and Wan~3.0~Prime share an identical tier-blind \texttt{ReservedCode1} prefix. %

\paragraph{Preliminary slices.}
Several secondary slices have small sample sizes and should be read as preliminary: \prelim{the nine single-clip \emph{Census} endpoints ($n = 1$), the cross-vendor survival matrix ($n \le 3$ per vendor and transform), the 480p substitution slice ($n = 9$), the second-session provenance checks ($n = 2$ sessions), and the 27 format-matched cross-session DPM comparisons (including three prompt pairs per endpoint on the six short-term replay endpoints)}. %
Seeded replay tests likewise cover three pairs per endpoint over a single $\sim$3~h window. %

\paragraph{Untested items.}
File-carried signatures attest to vendor claims rather than model weights, denoising steps, or quantisation. %
We did not evaluate provider cure following original-file demands, vendor-direct seed determinism, passive visual detection of unsigned S3 compute downgrades, or frame-deduplication mediation in zero-shot detectors, nor did we observe live substitution or estimate its prevalence in the wild. %

%% file: sections/07_conclusion.tex
\section{Conclusion}\label{sec:conclusion}

A video file returned by a commercial API establishes far less on its own about its generating model and tier than raw benchmark scores might suggest.
Six-way endpoint accuracy fell after uniform re-encoding, which also reduced the available features to bitstream statistics. Changing the effective frame rate of the same real clips shifted the two zero-shot detectors in opposite directions and reversed their post-hoc ranking despite a constant nominal frame rate. %
Even after uniform re-encoding, pairwise pixel separability across 45 model pairs varies substantially and drops to near chance between tiers of some models, while unverified gateway labels leave unsigned tier comparisons open to circularity. %
For temporal detectors, we recommend reporting the effective unique-frame rates of both real and generated cohorts and testing matched rates. For endpoint attribution, reporting results on both raw and uniformly re-encoded deliveries, with the feature set at each stage, makes the sensitivity to processing visible. %

Deployed provenance closes only part of this evidentiary gap today.
Pinned-policy C2PA verification authenticates tier claims for just one surveyed vendor, fails once routine remuxing or re-encoding strips the manifest (\prelim{in preliminary checks at $n \le 3$ per vendor}), and lacks request binding and trusted signing timestamps in the inspected Seedance store. %
A buyer can prove a signed tier contradiction against the order. A duplicate finding also requires a signed instance ID from an earlier authenticated, separately billed fresh-generation job, with retries and idempotent returns excluded. Unverifiable deliveries from signed endpoints warrant a demand for the vendor original under pass-through terms. Archive and stream-parameter matches elsewhere can justify review. %
Vendors could strengthen these checks by signing explicit tier identifiers and client request nonces and by obtaining trusted timestamps. Such evidence would link a vendor claim more closely to an individual order. Internal execution would still depend on trust in the signer. %

%% file: sections/09_appendix.tex
\section{Implementation Details}\label{sec:appendix}

\subsection{Review-Screen Construction and Diagnostics (ARM and DPM)}\label{app:arm}

\paragraph{Archive reuse matching (ARM).}
Given a delivered video, ARM decodes frame indices $0, 7, 14, \dots, 112$ (up to $m \le 17$ sampled frames) as $32 \times 32$ greyscale thumbnails. %
Spatial structure is captured by 63 AC DCT bits from the $8 \times 8$ low-frequency DCT block of the middle sampled frame, thresholded at their median. %
Temporal evolution is captured by $64m$ bits from stride-4 $8 \times 8$ sub-thumbnails across the $m$ sampled frames, thresholded against a \emph{single} mean pooled over all $m$ thumbnails, forming a $63 + 64m$-bit hash (1,151 bits at $m = 17$). %
Hashes of unequal length ($m_a \neq m_b$) receive normalised Hamming distance~1 and cannot match. %
Coarse SSIM samples latest frames at 8~fps, resizes to 64 pixels wide via area interpolation, and averages an $11 \times 11$ Gaussian window ($\sigma = 1.5$) over the common-duration frame prefix. %
ARM raises $\mathrm{RV}$ if and only if normalised Hamming distance is $\le 0.05$ and coarse SSIM is $\ge 0.90$. %
In initial 5-clip and 300-pair calibration checks, all 5/5 planted 480p CRF-28 copies matched (maximum hash distance 0.01825, coarse SSIM $\ge 0.9977$), whereas 0/300 random pairs matched and one hash-only collision at distance 0.0426 was rejected by coarse SSIM (0.3403). %

\begin{table}[t]
    \centering
    \caption{\textbf{ARM positive sensitivity and evasion diagnostics across 30 source clips.} Non-comparable pairs (unequal hash lengths $m$) are counted as misses.} %
    \label{tab:arm_perturbations}
    \footnotesize
    \setlength{\tabcolsep}{3pt}
    \adjustbox{max width=\linewidth}{%
\begin{tabular}{@{}lccc@{}}
        \toprule
        Perturbation / evasion transform & Matched & Non-comp. & Status \\
        \midrule
        H.264 re-encode, CRF~23 & 30/30 & 0 & Covered \\ %
        H.264 re-encode, CRF~28 & 30/30 & 0 & Covered \\ %
        Downscale to 480p (\texttt{p480}) & 29/30 & 0 & Covered \\ %
        Mild colour/contrast shift (\texttt{eq1}) & 29/30 & 0 & Covered \\ %
        Stronger colour/contrast shift (\texttt{eq2}) & 28/30 & 0 & Covered \\ %
        5\% playback speed-up (\texttt{speed105}) & 17/30 & 0 & Bypassed \\ %
        5\% spatial crop (\texttt{crop95}) & 3/30 & 0 & Bypassed \\ %
        10\% playback speed-up (\texttt{speed110}) & 1/30 & 25 & Bypassed \\ %
        10\% spatial crop (\texttt{crop90}) & 0/30 & 0 & Bypassed \\ %
        Horizontal flip (\texttt{hflip}) & 0/30 & 0 & Bypassed \\ %
        0.5\,s temporal trim (\texttt{trim05}) & 0/30 & 25 & Bypassed \\ %
        1.0\,s temporal trim (\texttt{trim10}) & 0/30 & 25 & Bypassed \\ %
        \bottomrule
    \end{tabular}%
}
\end{table}

Across all 586 originals in \emph{Main} (520) and \emph{Low-res} (66), spanning $\binom{586}{2} = 171{,}405$ pairs in total, ARM logged zero false matches: 0/168,081 on cross-prompt pairs (133,359 comparable and 34,722 non-comparable), 0/48 on same-model regenerations in \emph{Low-res} (24 different-seed, 18 replays of byte-identical requests, and 6 perturbed-prompt pairs; all 48 comparable), and 0/3,276 on other same-prompt cross-model pairs (2,245 comparable and 1,031 non-comparable), while matching 6/6 prompt-N02 480p re-encodes. %
Table~\ref{tab:arm_perturbations} details ARM's 30-source perturbation matrix: lossy re-encoding, 480p downscaling, and colour shifts remain well covered (28/30--30/30), whereas spatial crops, temporal trims, playback speed changes, and horizontal flips evade the fixed grid. %

\paragraph{Delivery-parameter matching (DPM).}
For each clip, DPM extracts nine stream fields via FFprobe: \texttt{width}, \texttt{height}, nominal frame rate \texttt{r\_frame\_rate}, packet-counted video frames \texttt{nb\_read\_packets}, audio presence, audio codec, audio sample rate, audio channel count, and audio-minus-video duration $d$. %
For each 720p delivery, DPM estimates the most frequent reference tuple(s) from genuine deliveries of the claimed endpoint in the other four prompt folds (48 reference clips in a 60-clip pool and 32 in a 40-clip pool). In the \prelim{nine-clip 480p Seedance~2.5 pool}, it excludes the delivery's base prompt and uses the other two base prompts (5--7 reference clips). %
A delivery passes if $|d - d_{\text{ref}}| \le 500\,\mu\text{s}$ and the remaining eight fields match exactly; otherwise DPM raises $\mathrm{RV}$. %

\subsection{Pinned-Policy C2PA Verifier, Adversarial Controls, and Survival Matrix}\label{app:verifier}

\paragraph{Strict verifier policy.}
Our strict verifier $F$ parses each file's C2PA manifest store using the C2PA reference reader, requires a \texttt{Valid} or \texttt{Trusted} validation state and both \texttt{claimSignature.validated} and \texttt{assertion.bmffHash.match}, and rejects every SDK failure code except \texttt{signingCredential.untrusted}. %
That exception is accepted only when independent pinned-path and validation-time checks pass; by contrast, an earlier relaxed check on 173 Seedance manifests ignored \texttt{signingCredential.untrusted} without enforcing a pinned signer policy. %
For BytePlus, $F$ pins the root CA's SHA-256 fingerprint and the leaf subject's \texttt{organization} and \texttt{organizationIdentifier}, and checks the certificate chain at the signed action time (\texttt{when}). For Black Forest Labs and Runway it checks the leaf organisation and requires a root in the frozen C2PA or Mozilla lists. Both policies use the signed action time when available and otherwise use the timestamp authority's time. Neither includes an OCSP query. %
The BytePlus audit issues one OCSP query for the shared leaf at audit time, failing the verifier audit if revoked and otherwise recording the status (verified \texttt{GOOD} with 0 revocations across BytePlus Seedance~2.x, whose three-certificate chain anchors at \emph{GlobalSign Secure Mail Root R45}: Mozilla e-mail trust bit \texttt{True}, C2PA trust list \texttt{False}). %

For BytePlus Level~3 claim consistency, a different known tier supports a contradiction. A missing, conflicting, or unknown label yields an unverifiable file. $F$ matches the signed \texttt{c2pa.actions.v2} tier string (\nolinkurl{dreamina-seedance-2-5}, \nolinkurl{dreamina-seedance-2-0-fast}, \nolinkurl{dreamina-seedance-2-0-mini}, and \prelim{\nolinkurl{dreamina-seedance-2-0} at $n = 1$}) against the ordered endpoint and checks the manifest instance ID against the ledger of authenticated billed jobs (192 authenticated billed IDs registered; 418 unauthenticated files excluded from the ID-history scan; 3 authenticated files not registered; 0 unauthenticated-history collisions). %
On all 187/187 audited Seedance clips, the signer-asserted \texttt{when} timestamp fell within the request submission-to-final-poll window widened by 60~s on each side; because \texttt{when} is signer-asserted rather than an RFC~3161 trusted timestamp, $F$ uses it only as the reference time for certificate validity and logs the window check diagnostically without gating on it. %

\begin{table}[t]
    \centering
    \caption{\textbf{Adversarial control validation of the strict C2PA verifier $F$} (18/18 controls placed in the expected class with the expected reason).} %
    \label{tab:verifier_controls}
    \footnotesize
    \setlength{\tabcolsep}{3pt}
    \adjustbox{max width=\linewidth}{%
\begin{tabular}{@{}lccl@{}}
        \toprule
        Control class & Count & $F$ output & Expected reason(s) \\
        \midrule
        1-byte \texttt{mdat} payload flip & 3/3 & \texttt{unverifiable} & \texttt{binding} \\ %
        Cross-tier manifest transplant & 3/3 & \texttt{unverifiable} & \texttt{binding} \\ %
        Forged signer chains (3 constructions) & 3/3 & \texttt{unverifiable} & \texttt{path}, \texttt{signature} \\ %
        Unmodified cheaper-tier swap & 3/3 & \texttt{contradictory} & \texttt{label\_mismatch} \\ %
        H.264 CRF-23 re-encode & 3/3 & \texttt{unverifiable} & \texttt{no\_manifest} \\ %
        Re-presented signed original & 3/3 & \texttt{duplicate} & \texttt{instance\_id\_seen} \\ %
        \bottomrule
    \end{tabular}%
}
\end{table}

\begin{table*}[t]
    \centering
    \caption{\textbf{Preliminary metadata presence and C2PA authentication survival matrix} (\prelim{preliminary, $n \le 3$ clips per vendor and transform}). ``Tag copy'' passes \texttt{-movflags use\_metadata\_tags} (plus \texttt{-map\_metadata 0} on CRF~23).} %
    \label{tab:survival_matrix}
    \footnotesize
    \setlength{\tabcolsep}{2.3pt}
    \adjustbox{max width=\linewidth}{%
\begin{tabular}{@{}lcccc@{}}
        \toprule
        Vendor group (regime; clips) & Remux / strip & Remux + tag copy & CRF~23 / 28 & CRF~23 + tag copy \\ %
        \midrule
        \prelim{Seedance ($n=3$), FLUX ($n=1$): C2PA auth.} & \prelim{0/3, 0/1} & \prelim{0/3, 0/1} & \prelim{0/3, 0/1} & \prelim{0/3, 0/1} \\ %
        \prelim{Seedance, Grok ($n=3$ each); FLUX, Runway ($n=1$ each): field} & \prelim{0/3 each; 0/1 each} & \prelim{0/3 each; 0/1 each} & \prelim{0/3 each; 0/1 each} & \prelim{0/3 each; 0/1 each} \\ %
        \prelim{Alibaba, Kuaishou, MiniMax ($n=3$ each): GB field} & \prelim{0/3 each} & \prelim{3/3 each} & \prelim{0/3 each} & \prelim{3/3 each} \\ %
        \bottomrule
    \end{tabular}%
}
\end{table*}

Tables~\ref{tab:verifier_controls} and~\ref{tab:survival_matrix} summarise the 18 adversarial controls across six attack classes and \prelim{the preliminary cross-vendor survival matrix across six remuxing and re-encoding transforms ($n \le 3$ per cell)}. %
Every tested FFmpeg remux and re-encode removed C2PA manifests across all four C2PA vendors (0 present, 0 authenticated), whereas GB~45438 user-data tags disappeared by default and survived only when explicit metadata-tag copying was requested. %

\subsection{Audit-Simulation Procedure and Scenario-Group Outcomes}\label{app:audit_simulation}

\paragraph{Resampling and bootstrap procedure.}
Our chain-level Monte Carlo simulation spans all 40 multi-prompt 720p fresh-substitution scenarios in \emph{Substitution}: three signed-endpoint pairs (Seedance~2.5 $\leftarrow$ 2.0~Mini, 2.0~Fast, and HappyHorse~1.1; $7 \times 3 = 21$ scenarios) and three unsigned pairs (Grok~1.5 $\leftarrow$ 1.5~Lite across 5 variants, plus Wan~3.0~Prime $\leftarrow$ Wan~3.0 and Kling~3.0~Std $\leftarrow$ Wan~3.0 across 7 variants each; 19 scenarios). %
For each scenario, audit configuration ($F$ alone, $F + \text{ARM}$, and $F + \text{ARM} + \text{DPM}$), and mixing rate $p \in \{0, 5\%, 10\%, 20\%\}$, we draw 10,000 simulated 30-slot audits using fixed seed \texttt{20261116}. %
Each slot independently samples a substitute clip with probability $p$ or an honest reference clip with probability $1 - p$ from that scenario's frozen evaluation pool (drawing from the collected corpus rather than issuing new API calls). %
We compute 95\% confidence intervals via 200 prompt-cluster bootstrap resamples of 2,000 simulated audits each, alongside the exact binomial reference in Eq.~\eqref{eq:power} for independent slots with $q = 0$ and $d \in \{0, 1\}$. %

\begin{table*}[t]
    \centering
    \caption{\textbf{Simulated 30-clip audit verdict probabilities across all 40 multi-prompt fresh-substitution scenarios} under the full graded chain ($F + \text{ARM} + \text{DPM}$; 10,000 resampled audits per scenario, min--max across scenarios in each group). At $p = 0$, every scenario and arm had 0/10,000 flags.} %
    \label{tab:sim_groups}
    \footnotesize
    \setlength{\tabcolsep}{3.2pt}
    \adjustbox{max width=\linewidth}{%
\begin{tabular}{@{}lcrrrrrr@{}}
        \toprule
        Scenario group (count) & $p$ & Accuse ($\mathrm{AC}$) & Auto.\ breach ($\ge 4/30$) & Pending demand & Review ($\mathrm{RV}$) & Abstain / $\mathrm{CC}$ & Review+ \\ %
        \midrule
        Signed tier swap, raw (2) & 0.05 & 0.7850--0.7860 & 0.0000--0.0000 & 0.0000--0.0000 & 0.0000--0.0000 & 0.2140--0.2150 & 0.7850--0.7860 \\ %
        Signed tier swap, raw (2) & 0.10 & 0.9563--0.9587 & 0.0000--0.0000 & 0.0000--0.0000 & 0.0000--0.0000 & 0.0413--0.0437 & 0.9563--0.9587 \\ %
        Signed tier swap, raw (2) & 0.20 & 0.9988--0.9991 & 0.0000--0.0000 & 0.0000--0.0000 & 0.0000--0.0000 & 0.0009--0.0012 & 0.9988--0.9991 \\ %
        \midrule
        Signed, proc.\ or cross-vendor (19) & 0.05 & 0.0000--0.0000 & 0.0601--0.0627 & 0.7233--0.7249 & 0.0000--0.0000 & 0.2140--0.2150 & 0.7850--0.7860 \\ %
        Signed, proc.\ or cross-vendor (19) & 0.10 & 0.0000--0.0000 & 0.3476--0.3524 & 0.6063--0.6087 & 0.0000--0.0000 & 0.0413--0.0437 & 0.9563--0.9587 \\ %
        Signed, proc.\ or cross-vendor (19) & 0.20 & 0.0000--0.0000 & 0.8718--0.8751 & 0.1240--0.1270 & 0.0000--0.0000 & 0.0009--0.0012 & 0.9988--0.9991 \\ %
        \midrule
        Unsigned, unnorm.\ diff.\ tuple (12) & 0.05 & 0.0000--0.0000 & 0.0000--0.0000 & 0.0000--0.0000 & 0.7850--0.7860 & 0.2140--0.2150 & 0.7850--0.7860 \\ %
        Unsigned, unnorm.\ diff.\ tuple (12) & 0.10 & 0.0000--0.0000 & 0.0000--0.0000 & 0.0000--0.0000 & 0.9563--0.9587 & 0.0413--0.0437 & 0.9563--0.9587 \\ %
        Unsigned, unnorm.\ diff.\ tuple (12) & 0.20 & 0.0000--0.0000 & 0.0000--0.0000 & 0.0000--0.0000 & 0.9988--0.9991 & 0.0009--0.0012 & 0.9988--0.9991 \\ %
        \midrule
        Unsigned, shared tuple (7) & 0.05 & 0.0000--0.0000 & 0.0000--0.0000 & 0.0000--0.0000 & 0.0000--0.0000 & 1.0000--1.0000 & 0.0000--0.0000 \\ %
        Unsigned, shared tuple (7) & 0.10 & 0.0000--0.0000 & 0.0000--0.0000 & 0.0000--0.0000 & 0.0000--0.0000 & 1.0000--1.0000 & 0.0000--0.0000 \\ %
        Unsigned, shared tuple (7) & 0.20 & 0.0000--0.0000 & 0.0000--0.0000 & 0.0000--0.0000 & 0.0000--0.0000 & 1.0000--1.0000 & 0.0000--0.0000 \\ %
        \bottomrule
    \end{tabular}%
}
\end{table*}

\paragraph{Honest gate and ablation contrasts.}
At $p = 0$, every scenario and all three audit configurations produced zero accusations, automatic breaches, original-file demands, or review flags across all 10,000 resampled audits. %
All 187 Seedance deliveries passed the strict check, and all 173 files in the earlier relaxed check passed that check. With zero failures, the illustrative one-sided 95\% binomial upper bounds are 1.59\% and 1.72\%, respectively. Using the latter as an independent per-clip failure probability gives a 0.17\% chance of at least four failures in 30 deliveries. Clips share prompts, so these figures do not establish a deployment false-breach rate. %
Table~\ref{tab:sim_groups} reports the group-level probabilities across $p \in \{5\%, 10\%, 20\%\}$; on the 12 unsigned tuple-distinct scenarios at $p = 10\%$, 95\% prompt-cluster bootstrap confidence intervals on Review+ span lower bounds of 0.9495--0.9500 to upper bounds of 0.9665--0.9675. %
Across all 40 fresh-substitution scenarios, $F + \text{ARM}$ matched $F$ identically (0.0000 maximum difference), whereas adding unnormalised DPM increased $p = 20\%$ Review+ by $+0.9988$ ($[0.997, 1.000]$) on Grok~1.5 $\leftarrow$ 1.5~Lite and $+0.9991$ ($[0.997, 1.000]$) on Kling~3.0~Std $\leftarrow$ Wan~3.0, with zero change elsewhere. %